\documentclass[pre,twocolumn,amsmath,amssymb,nofootinbib,floatfix,superscriptaddress]{revtex4}

\usepackage{graphicx,bm,xcolor, bbold}
\usepackage{enumerate}
\usepackage{graphicx,bm}
\usepackage[normalem]{ulem}

\makeatletter
\def\graphicscale{\twocolumn@sw{0.3}{0.4}}
\def\graphicthreescale{\twocolumn@sw{0.3}{0.4}}

\begin{document}

\title{Finite-temperature topological transitions in the presence of\\ quenched
  uncorrelated disorder}

\author{Claudio Bonati} \affiliation{Dipartimento di Fisica
  dell'Universit\`a di Pisa and INFN Sezione di Pisa, Largo Pontecorvo 3,
  I-56127 Pisa,
  Italy}

\author{Ettore Vicari} 
\affiliation{Dipartimento di Fisica dell'Universit\`a di Pisa,
        Largo Pontecorvo 3, I-56127 Pisa, Italy}

\date{\today}

\begin{abstract}
We address issues related to the presence of defects at
finite-temperature topological transitions, in particular when
defects are modeled in terms of further variables associated with a
quenched disorder, corresponding to the limit in which the defect
dynamics is very slow. As a paradigmatic model, we consider the {\em
  classical} three-dimensional lattice ${\mathbb Z}_2$ gauge model in
the presence of quenched uncorrelated disorder associated with the
plaquettes of the lattice, whose topological transitions are
characterized by the absence of a local order parameter. We study the
critical behaviors in the presence of weak disorder. We show that they
belong to a new topological universality class, different from that of
the lattice ${\mathbb Z}_2$ gauge models without disorder, in
agreement with the Harris criterium for the relevance of uncorrelated
quenched disorder when the pure system undergoes a continuous
transition with positive specific-heat critical exponent.
\end{abstract}

\maketitle

\section{Introduction}
\label{intro}

Statistical systems with quenched disorder, modeling the presence of
defects subject to a slow dynamics, generally develops peculiar
critical phenomena associated with new universality classes with
respect to pure systems, such as transitions to glassy
phases. Finite-temperature phase transitions and critical behaviors in
disordered systems have been mostly studied within lattice spin models
with various realizations of local disorder, in particular with types
of disorder preserving the symmetry of the pure system, such as
random-bond and random-site disorder (random-field disorder instead
generally breaks the global symmetry), and spatially uncorrelated
probability distributions of the quenched variables, see, e.g.,
Refs.~\cite{Ma-book, Nishimori-book, Cardy-book, Harris-74, EA-75, Nishimori-81,
  HPPV-07, HPPV-07-rdi, HPPV-07-mc, PC-99, Betal-00, KM-02, KR-03,
  KKY-06, Jorg-06, HPV-08, BFMMMY-11, Janus-13, LPP-16, Granato-04,
  MHT-04, PR-05, LY-07, VK-09, FMPTY-09, BFMS-14, OUK-20, AV-11,
  Asp-etal-16, Hartmann-99, CPV-11, PHP-06, LH-88, HPPV-08, PPV-09,
  FMMPR-16, CP-19, Nishimori-24,Vojta-19}.

While a large amount of disorder may lead to glassy phases and glassy
critical behaviors in {\em classical} three-dimensional (3D) spin
systems, ferromagnetic transitions may still occur for sufficiently
weak disorder.  However, even though the quenched disorder does not
change the symmetry-breaking pattern, the universality class of the
critical behavior may differ from that of the ferromagnetic transition
of the pure system. According to the Harris criterium~\cite{Harris-74}
[see, e.g., \cite{Ma-book, Cardy-book} for the
    renormalization-group (RG) interpretation/generalization of the
    original argument], which applies to a spatially-uncorrelated
quenched disorder that is effectively coupled to the energy density,
transitions in the presence of weak disorder may develop a new
universality class when the specific-heat exponent $\alpha$ of the
pure transition is positive, as it happens in 3D random-bond and
random-site Ising
systems~\cite{Ma-book,Nishimori-book,HPPV-07,HPPV-07-rdi}. On the
other hand, if the pure-system continuous transition has a negative
specific-heat exponent, then its universality class turns out to be
stable with respect to weak disorder, as it happens in 3D random-bond
O($N$) vector models for any $N\ge 2$, such as the XY and Heisenberg
spin models~\cite{PV-02}, corresponding to $N=2$ and $N=3$
respectively.  Moreover, with increasing disorder, spin systems may
develop peculiar multicritical behaviors along the so-called Nishimori
line~\cite{Nishimori-book,Nishimori-81,LH-88,HPPV-07-mc,HPPV-08,AV-11},
and then transitions to glassy phases, see, e.g.,
Refs.~\cite{KKY-06,HPV-08,BFMMMY-11,Janus-13,LPP-16,Granato-04,MHT-04,PR-05,
  LY-07,VK-09,FMPTY-09,BFMS-14,OUK-20,AV-11,Asp-etal-16,CPV-11}.

The effects of quenched disorder have been much less investigated at
topological transitions, such as those occurring in systems with gauge
symmetries, without a local order parameter driving the
transition~\cite{Wegner-71,Kogut-79,Sachdev-19,BPV-25}.  A
paradigmatic model for systems undergoing finite-temperature
topological transitions is provided by the 3D lattice ${\mathbb Z}_2$
gauge model~\cite{Wegner-71,Kogut-79}, whose confinement-deconfinement
transition is not driven by a local order parameter, but by the
nonlocal order parameter associated with the area/perimeter law of the
Wilson loops~\cite{Wegner-71,Kogut-79}. 3D lattice ${\mathbb Z}_2$
gauge models have been also considered in the presence of local
spatially-uncorrelated quenched disorder associated with each
plaquette of the lattice~\cite{DKLP-02,WHP-03} (denoted by RPGM, for
random plaquette gauge model, in the following), i.e., with randomly
chosen plaquettes having couplings with the {\em wrong} (negative
instead of positive) sign.  The RPGM was originally introduced in
relation to the theory of the quantum error correction. However, it
also provides a paradigmatic example of systems undergoing topological
transitions in the presence of disorder. Some results on its
temperature-disorder phase diagram were reported in
Refs.~\cite{WHP-03,OAIM-04}, but the critical behavior at the
transition lines has never been analyzed. Therefore the
characterization of the universality classes of its topological
transitions in the presence of quenched disorder is still an open
issue.

In this paper we investigate the effects of disorder at
finite-temperature topological transitions, focusing on the
paradigmatic 3D RPGM.  We recall that the pure 3D lattice ${\mathbb
  Z}_2$ gauge model is related to the standard 3D Ising model by a
duality relation~\cite{Wegner-71,Savit-80}, thus they share the same
length-scaling critical exponent $\nu_{\cal I}\approx
0.630$~\cite{PV-02}, which implies that the specific-heat exponent is
positive, $\alpha_{\cal I}\approx 0.110$. We present a numerical
analysis of the critical behaviors of the 3D lattice ${\mathbb Z}_2$
gauge model for weak disorder, which is made difficult by the fact
that its topological transitions do not have gauge-invariant local
order parameters.  We thus study the critical behaviors of this system
by focusing on the scaling behavior of the gauge-invariant energy
cumulants. The numerical results show that the quenched disorder
destabilizes the pure-system topological transition.  Indeed, we
observe critical behaviors belonging to another universality class
with $\nu = 0.82(2)$, which is significantly larger that the Ising
length-scale exponent $\nu_{\cal I}\approx 0.630$ in the absence of
disorder (and also different from any universality class of Ising
systems with quenched disorder).  In the final part of the paper we
also briefly discuss generalizations of the RPGM, considering generic
3D lattice ${\mathbb Z}_N$ gauge models with weak disorder.

While in this work we focus on the effect of weak disorder at
finite-temperature topological transitions, it is important to note
that the presence of disorder can lead to very interesting
phenomenological consequences also in many quantum (i.e., zero
temperature) contexts, in particular for what regards quantum critical
phenomena, see, e.g., Refs.~\cite{Sachdev-book-qpt, Vojta-19}, and
topological phases of matter, see, e.g.,
Refs.~\cite{MoessnerMoore-book, Sachdev-book-phases, SM-09, GWATB-09,
  GRRF-10, SLJSX-12, GLR-13, TLRR-15, OSBS-16, AFL-24, SCMVL-24,
  ZT-26}.

The paper is organized as follows: in Sec.~\ref{model} we introduce
the RPGM and discuss some properties of its phase diagram, then in
Sec.~\ref{fss} we investigate the critical behavior of this model in
the regime of weak disorder, providing a quite accurate estimate of
the corresponding critical exponent $\nu$.  Finally, in
Sec.~\ref{conclu} we report our conclusions and discuss some
generalizations of the RPGM.

\section{The model}
\label{model}

In the statistical RPGM, i.e., a 3D lattice ${\mathbb Z}_2$ gauge
model with uncorrelated quenched disorder associated with the
plaquettes, spin variables $\sigma_{{\bm x},\mu}=\pm 1$ are associated
with the links (starting from site ${\bm x}$ in the positive $\mu$
direction, $\mu=1,2,3$) of a cubic lattice of size $L$, and quenched
disorder variables $w_{{\bm x},\mu\nu}=\pm 1$ are associated with the
plaquettes identified by the site ${\bm x}$ and the directions
$\mu>\nu$. Its Hamiltonian reads~\cite{Wegner-71,DKLP-02,WHP-03}
\begin{eqnarray}
  H=-K\,\mathcal{H}\ ,\quad \mathcal{H} = \sum_{{\bm x},\mu>\nu}
  w_{{\bm x},\mu\nu} \, \Pi_{{\bm x},\mu\nu}, \label{Hamdz2g}
\label{eqnarray}
\end{eqnarray}
where $K$ is the gauge coupling, $\Pi_{{\bm x},\mu\nu}$ with $\mu>\nu$
is the plaquette operator
\begin{eqnarray}
\Pi_{{\bm x},\mu\nu}=
  \sigma_{{\bm x},\mu} \,\sigma_{{\bm x}+\hat{\mu},\nu} \,\sigma_{{\bm
      x}+\hat{\nu},\mu} \,\sigma_{{\bm x},\nu},
\label{plaquette}
\end{eqnarray}
$\hat\mu$ indicates the unit vector along the direction $\mu$ and
periodic boundary conditions are used on the lattice. The quenched
disorder variables $w_{{\bm x},\mu\nu}$ ($\mu>\nu$) are spatially
uncorrelated, and each disorder configuration is chosen according to
the probability distribution
\begin{equation}
  P_w(q) = \prod_{{\bm x},\mu>\nu} \Bigl[ (1-q) \,\delta(w_{{\bm x},\mu\nu} -
      1) + q \, \delta(w_{{\bm x},\mu\nu} + 1)\Bigr],
\label{probdis}
\end{equation}
where $0\le q \le 1$ is a global parameter. Therefore, $q$ is the
probability of getting an extra minus sign in the Hamiltonian weight
of the plaquette.  The free-energy density of the RPGM is obtained by
averaging over disorder configurations:
\begin{eqnarray}
&& F(K) = \frac{1}{V} \sum_{\{w\}} P_w(q) \, \ln
  Z_w(K), 
  \label{Fkp}\\ && Z_w(K) = \sum_{\{\sigma\}} e^{-H/T},
  \qquad V=L^3,
\nonumber 
\end{eqnarray}
where $\{w\}$ and $\{\sigma\}$ indicate disorder and spin
configurations respectively.  Without loss of generality, in the
following we fix $T=1$. Therefore, the relevant phase diagram of the
lattice ${\mathbb Z}_2$ gauge model with uniform spatially
uncorrelated quenched disorder should be studied in the $q$-$K$ plane.
  
In the absence of disorder, which corresponds to the model in
Eq.~(\ref{Hamdz2g}) for $q=0$, the 3D lattice ${\mathbb Z}_2$-gauge
model undergoes a continuous topological transition at a finite value
of $K$~\cite{Wegner-71,Kogut-79,Sachdev-19,BPV-25}, $K_c(q=0)=
0.761413292(11)$, separating the high-$K$ deconfined phase from the
low-$K$ confined phase.  This transition is topological, as it is not
driven by a local order parameter. The behavior of the nonlocal Wilson
loop $W_C$, defined as the product of the link variables along a
closed contour $C$ within a plane, provides a nonlocal order parameter
for the transition~\cite{Wegner-71,Kogut-79}. Indeed, its size
dependence for large contours changes at $K_c$, from an area law
$W_C\sim \exp(- c_a A_C)$ for small values of $K$, where $A_C$ is the
area enclosed by the contour $C$ and $c_a>0$ is the so called string
tension, to a perimeter law $W_C\sim \exp(- c_p P_C)$ for large values
of $K$, where $P_C$ is the perimeter of the contour $C$ and $c_p>0$ is
a constant.  The topological transition of the 3D lattice ${\mathbb
  Z}_2$ gauge model can be related to the continuous transition of the
standard 3D Ising model, because a nonlocal duality mapping relates
the partition functions of the two
models~\cite{Wegner-71,Savit-80}. Duality implies that thermal
observables have the same critical behavior in the two models and, in
particular, the correlation-length exponent $\nu$ is the same.  The
most accurate estimate for the 3D Ising universality class is
$\nu_{\cal I}=0.629971(4)$~\cite{KPSV-16}, see also
Refs.~\cite{PV-02,Hasenbusch-21,FXL-18,KP-17,Hasenbusch-10,CPRV-02,GZ-98}
for other results.~\footnote{The universal features of the topological
confinement-deconfinement transitions of 3D lattice ${\mathbb Z}_N$
models have been investigated in several works, see, e.g.,
Refs.~\cite{Wegner-71,BDI-74, BKKLS-90, CFGHP-97, GPP-02, BCCGPS-14,
  XCMCS-18, ACFIP-24,BPV-25,BPV-25-dyn,BPV-25-rel,BPV-25-zn}.}

\begin{figure}
  \includegraphics[width=0.8\columnwidth]{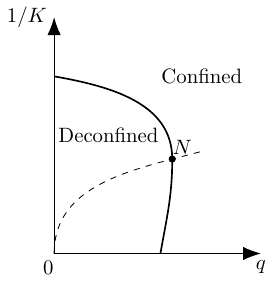}
  \caption{The expected phase diagram of the 3D RPGM, see
    Refs.~\cite{WHP-03,OAIM-04}. The continuous line separates the
    deconfined phase, at small $K^{-1}$ and $q$, from the confined
    phase, while the dashed line denotes the Nishimori line. The black
    dot $N$ represents the point where the deconfinement transition
    line intersects the Nishimori line, and the value of $q$ at this
    point is denoted by $q_N$ in the main text.}
  \label{phadia}
\end{figure}

The phase diagram for $q>0$ has been discussed in
Refs.~\cite{WHP-03,OAIM-04}.  A sketch of the $q$-$K^{-1}$ phase
diagram is shown in Fig.~\ref{phadia}.  It shows notable analogies
with the phase diagrams of the random-bond Ising and random-phase XY
models, see, e.g., Refs.~\cite{Nishimori-81, LH-88, Hartmann-99,
  HPPV-07-mc, HPV-08, CPV-11, PHP-06, HPPV-08, PPV-09, AV-11}.  The
deconfined phase at sufficiently large values of $K$ is expected to
extend to nonzero values of $q$, and it is separated from the confined
phase by a transition line, which starts from the critical point along
the $q=0$ line.  The so-called Nishimori line~\cite{Nishimori-book,
  Nishimori-81}, where $K$ and $q$ satisfy the relation~\cite{WHP-03}
\begin{equation}
  e^{-2K}  =   \frac{q}{1-q},
  \label{nishimoriline}
\end{equation}  
show an enhanced symmetry, which allows us to derive a number of exact
relations~\cite{WHP-03}.  In particular, the maximum value of $q$
showing a deconfined phase corresponds to the location $q_N$ of the
topological confining-deconfining transition along the Nishimori line.
Ref.~\cite{OAIM-04} reported the estimate $q_N\approx 0.033$.  The
transition line below the Nishimori critical point is likely slightly
reentrant, as it also occurs in random-bond Ising models, see, e.g.,
Ref.~\cite{PPV-09,CPV-11}.  There is no evidence of the existence of
further gauge glassy phase for $q>q_N$, as characterized in
Ref.~\cite{WHP-03}, although its presence cannot be excluded yet.
Therefore, the phase diagram should be similar to that of the 2D
random-bond Ising
model~\cite{Nishimori-81,PHP-06,PPV-09,FMMPR-16,WHP-03} with the
ferromagnetic and paramagnetic phases replaced by the deconfined and
confined phases, as sketched in Fig.~\ref{phadia}.  In this work we
focus on the critical properties at the confined-deconfined
topological transition for small disorder parameter $q$, up to the
critical point $N$ along the Nishimori line.

According to the Harris criterium~\cite{Harris-74} (see also, e.g.,
\cite{Ma-book, Cardy-book}), spatially uncorrelated quenched disorder coupled
to the energy density of the system provides a relevant perturbation at the
fixed points corresponding to the critical behavior of pure systems with
positive specific-heat exponent. This is for example the case of the 3D
random-bond Ising model where the specific-heat exponent of the pure
Ising transition is $\alpha_{\cal I}=2-3\nu_{\cal I}\approx 0.110$. In
this case the critical behavior at the ferromagnetic transitions in
the presence of spatially uncorrelated disorder belong to a different
universality class, the so called randomly-dilute Ising (RDI)
universality class, whose length-scale critical exponent is $\nu_{\rm
  rdi}=0.683(2)$~\cite{HPPV-07-rdi, HPPV-07, PV-00}, and the
corresponding specific-heat exponent is negative, $\alpha_{\rm rdi} =
2 - 3\nu_{\rm rdi} = -0.049(6)$.  An analogous change of universality
class occurs also in the 3D RPGM for $q>0$.  Indeed, also in this case
we have spatially uncorrelated quenched variables coupled to the
energy density, which corresponds to the plaquette term in lattice
${\mathbb Z}_2$ gauge models.  Therefore, the universality class of
the disordered topological transitions is expected to differ from that
of the pure model at $q=0$. Note also that there is no reason to
expect that such transition belongs to the RDI universality class of
disordered spin systems, since the duality mapping with the Ising
model is lost in the presence of disorder.

We finally mention that the results reported in Ref.~\cite{WHP-03,OAIM-04}
along the confining-deconfining transition do not give hints of the
universality class of the critical behavior for weak disorder, and in
particular of the corresponding critical exponent $\nu$.

\section{Numerical methods and results}
\label{fss}

Topological transitions are notoriously difficult to investigate, due
to the absence of local order parameters. It might seem that the
natural strategy for investigating the topological transitions of
lattice ${\mathbb Z}_2$ gauge models is to study the area or perimeter
law behavior of large Wilson loops, as done in \cite{OAIM-04}. Such a
strategy is however effective only if we want to investigate the phase
structure of the model, and it becomes more and more inefficient as we
approach a continuous phase transition.  It indeed requires the use of
very large lattices to estimate the (vanishing at the transition)
string tension from the large distance behavior of Wilson loops.  In
our numerical study we follow a more convenient numerical approach,
which lends itself for a finite-size scaling (FSS) analysis at the
critical point.

We study the FSS behavior of the gauge-invariant energy cumulants,
normalized by the lattice volume and averaged over disorder. These
quantities, which  will be denoted by $B_k$ in the following, are the
intensive quantities defined by 
\begin{equation}
\begin{aligned}
B_k&=\left(\frac{\partial}{\partial{K}}\right)^k F(K)=\\
&=\frac{1}{V}
\sum_{\{w\}} P_w(q) \, \left(\frac{\partial}{\partial{K}}\right)^k
\ln  Z_w(K).
\end{aligned}
\label{bkdef}
\end{equation}
The cumulants can be related (for $k>1$) to the central moments of the
energy defined for each disorder realization by 
\begin{equation}
  m_k = \langle \, (\mathcal{H}-\langle \mathcal{H}\rangle)^k \rangle,
  \label{mkmom}
\end{equation}
where $\mathcal{H}$ is defined in Eq.~\eqref{Hamdz2g}, and the
statistical average $\langle\phantom{a}\rangle$ is taken over the spin
variables at fixed disorder configuration. One can easily see that
$B_1$ is proportional to energy density, i.e.,
\begin{equation}
  B_1=\frac{1}{V}[\langle \mathcal{H}\rangle]_w,
  \label{b1h}
\end{equation}
where the square brackets $[\phantom{a}]_w$ indicate the average over
the quenched disorder, while the second and third cumulants are given
by
\begin{eqnarray}
B_k=\frac{1}{V} [m_k]_w \quad{\rm for}\;\;k=2,3.
\label{B23def}  
\end{eqnarray}
In particular, $B_2$ is proportional to the specific heat. The
relation between cumulants and central moments becomes more
complicated for higher cumulants, for example
\begin{equation}
B_4 = \frac{1}{V} [m_4 - 3 m_2^2]_w.
\label{B4def}
\end{equation}

The energy cumulants are very useful to characterize topological
transitions in which no local gauge-invariant order parameter is
present (see, e.g., Refs.~\cite{SSNHS-03, BPV-20-hcAH, BPV-22-z2g,
  BPV-24-coH, BPV-24}), since for fixed $q$ they are expected to show the FSS
behavior~\cite{SSNHS-03, BPV-24-coH}
\begin{eqnarray} 
&&B_k(K,L) \approx L^{k/\nu-3} {\cal B}_k(X) \left[1 + L^{-\omega}
    {\cal B}_{k,\omega}(X)\right] + b_k,
\nonumber\\
&&X = (K-K_c)L^{1/\nu},
\label{Hg3-scaling}
\end{eqnarray}
where the constants $b_k$ represent the analytic
background~\cite{PV-02,BPV-24-coH}, and the $O(L^{-\omega})$
suppressed term with $\omega>0$ represents the leading scaling
correction which is generally associated with the leading irrelevant
RG perturbation at the fixed point~\cite{PV-02} (for example
$\omega\approx 0.8$ at the standard Ising transition, i.e., for
$q=0$).  The scaling functions ${\cal B}_k(X)$ are universal apart
from a multiplicative factor and a normalization of the argument,
however they generally depend on the boundary conditions adopted.

It is important to stress that the background term $b_k$ in
Eq.~(\ref{Hg3-scaling}) turns out to be subleading with respect to the
scaling term only when $k-3\nu >0$, thus for $k\ge 2$ when the
specific-heat exponent is positive, i.e., $\alpha=2-3\nu>0$, which is
the case for $q=0$. If instead $\alpha<0$, which is the case expected
along the confining-deconfining transition line for $q>0$ due to
Harris criterium, higher cumulants are needed to identify the scaling
behavior, with minimum value of $k$ depending on the critical exponent
$\nu$. Note however that higher cumulants also become more noisy, so
it is typically convenient to use the cumulant corresponding to the
smallest value of $k$ such that $k-3\nu >0$.

\begin{figure}
  \includegraphics[width=0.95\columnwidth]{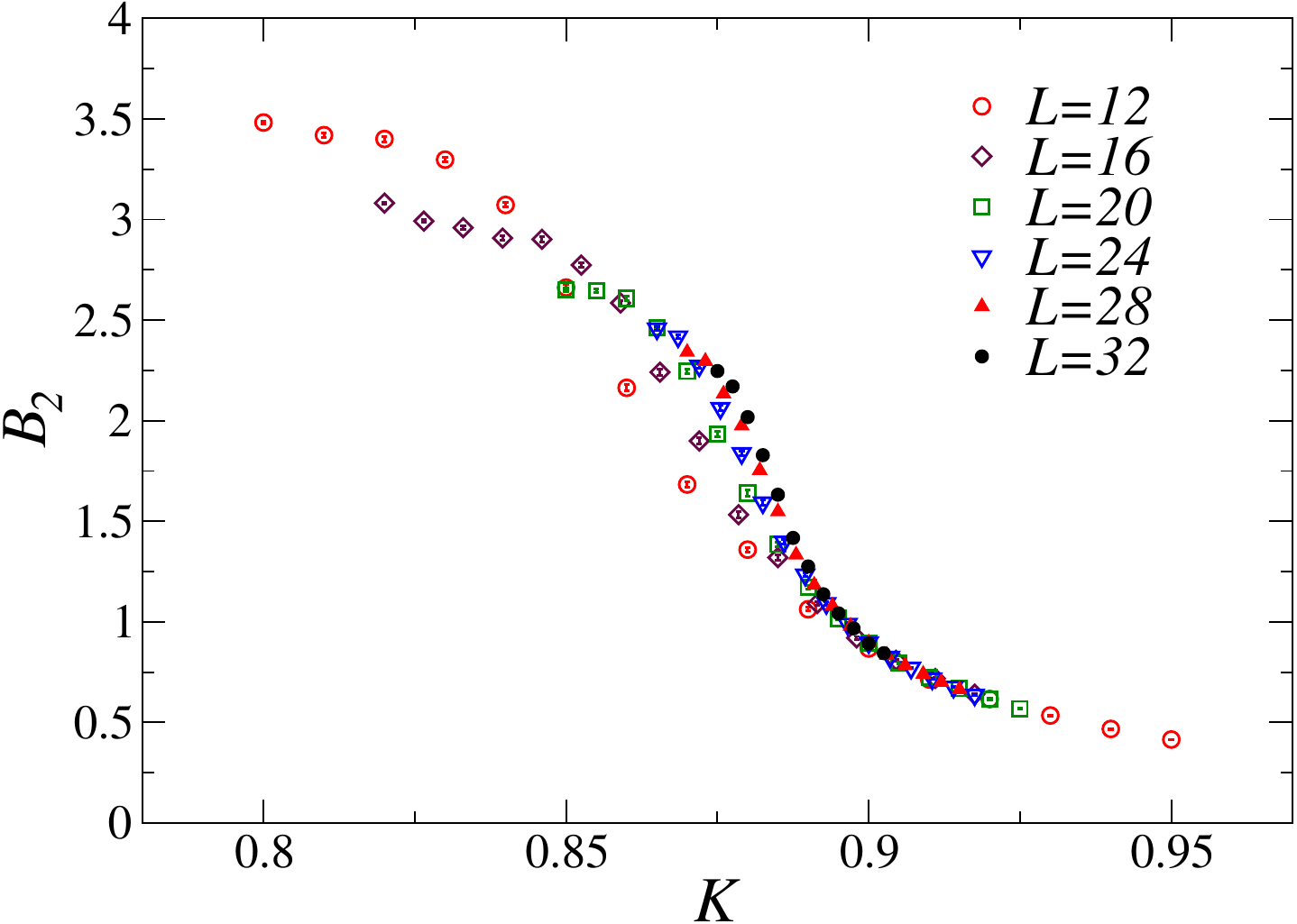}
  \caption{Data for the second cumulants $B_2$, proportional to the
    specific heat, across the topological transition for $q=0.015$.}
  \label{enespecheat}
\end{figure}

\begin{figure}
  \includegraphics[width=0.95\columnwidth]{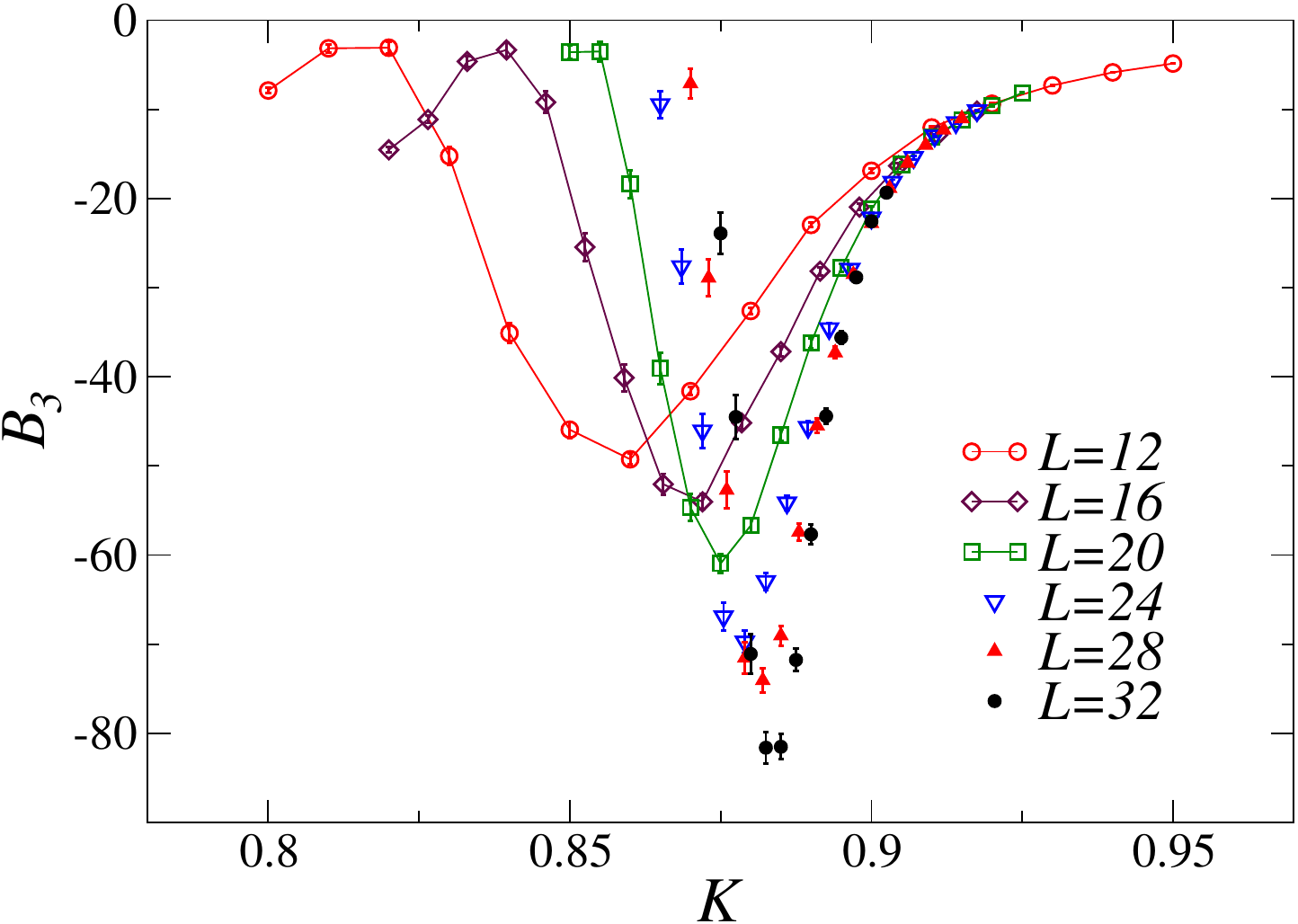}
  \caption{ Data for the third cumulant $B_3$ for $q=0.015$, across
    the topological transition around $K\approx 0.89$. Continuous
    lines for $L=12, 16, 20$ are drawn just to guide the eye.}
  \label{thirdcum}
\end{figure}

To characterize the critical behavior of the RPGM for weak disorder we
present a numerical FSS analysis for $q=0.015$, which according to the
results of Ref.~\cite{OAIM-04} should be sufficiently small for the
system to develop a finite-temperature confining-deconfining
transition. We thus perform simulations for $q=0.015$ and several
values of $K$, for lattice sizes up to $L=32$. We use $O(10^3)$
disorder samples and for each disorder realization
we carry out $2\times 10^5$ Metropolis updates of the whole lattice,
always starting from the ordered configuration and
discarding one quarter (half, for the larges lattices) of them as
thermalization.  Moreover, three independent simulations have been
carried out for each disorder realization, in order to
obtain unbiased estimates of the first three cumulants of the energy
(see \cite{HPPV-07-rdi} for a discussion of this point).

\begin{table}[t]
\begin{tabular}{llllll} 
\hline\hline
$L_{\mathrm{min}}$ & fit range & scal. corr. & $K_c$ & $\nu$ &  $\chi^2$/dof \\ \hline  
20 & [-1.5,1] & no & 0.89436(33) & 0.8162(56) & 69/42 \\
24 & [-1.5,1] & no & 0.89442(53) & 0.826(10) & 32/28 \\ \hline
20 & [-1,1] & no & 0.89393(50) & 0.8102(76) & 65/35 \\
24 & [-1,1] & no &  0.89396(82) & 0.821(14) & 27/21 \\ \hline
20 & [-0.5,0.5] & no & 0.8933(17) & 0.811(31) & 12/11 \\
24 & [-0.5,0.5] & no & 0.8919(23) & 0.782(46) & 12/9 \\ \hline
12 & [-1.5,1] & fit $\kappa$ & 0.89371(29) & 0.8129(54) & 67/58 \\
16 & [-1.5,1] & fit $\kappa$ & 0.89414(63) & 0.822(15) & 46/46 \\ \hline
16 & [-1.5,1] & $\kappa=0.5$ & 0.89389(42) & 0.802(10) & 70/55 \\ 
20 & [-1.5,1] & $\kappa=0.5$ & 0.89406(67) & 0.792(21) & 36/36 \\ \hline
16 & [-0.5,0.5] & $\kappa=0.5$ & 0.89370(98) & 0.827(14) & 41/28 \\
20 & [-0.5,0.5] & $\kappa=0.5$ & 0.8927(21) & 0.813(32) & 18/17 \\ \hline
16 & [-1.5,1] & $\kappa=1.0$ &  0.89382(40) & 0.804(10) & 69/55 \\
20 & [-1.5,1] & $\kappa=1.0$ &  0.89301(69) & 0.800(20) & 46/38 \\ \hline
16 & [-0.5,0.5] & $\kappa=1.0$ &  0.89298(91) & 0.817(13) & 32/28 \\
20 & [-0.5,0.5] & $\kappa=1.0$ &  0.8932(20) & 0.820(30) & 16/16 \\ \hline
\end{tabular}
\caption{Some of the fits performed to estimate $K_c$ and $\nu$ for
  $q=0.015$.  $L_{\mathrm{min}}$ denotes the smallest value of the
  lattice size considered in the fit, and the fit range refers to the
  values of $X$, cf. Eq.~\eqref{Hg3-scaling}, self-consistently used
  in the fit. Scaling corrections have been either totally neglected,
  or taken into account by fixing the reported value of $\kappa$ (see
  text), or taken into account account by leaving $\kappa$ as a free
  parameter in the fit.  In the last case, however, the value of
  $\kappa$ could not be reliably estimated due to large statistical
  uncertainties.  The errors on the various results are only
  statistical. }
\label{TabFitRes}
\end{table}

The numerical results obtained for the second cumulant $B_2$,
corresponding to the specific heat, are shown in
Fig.~\ref{enespecheat}.  They do not appear to diverge with increasing
$L$, implying that $\alpha<0$, consistently with the expectations
coming from Harris criterium.  In Fig.~\ref{thirdcum} we report data
for the third cumulant $B_3$, which is the first cumulant to display
an apparently divergent behavior at the transition. To estimate $K_c$
and $\nu$ we performed several fits using the FSS relation in
Eq.~\eqref{Hg3-scaling}, with a polynomial approximation of
$\mathcal{B}_k(X)$ and taking also into account scaling corrections,
i.e., using the functional form
\begin{equation}
  B_3(K,L) = L^{3/\nu-3}\left(\sum_{i=0}^{n_l} a_i X^i
  + L^{-\kappa}\sum_{i=0}^{n_s} b_i X^i\right).
\end{equation}
If $\omega < 3/\nu-3$ then $\kappa=\omega$ and the $O(L^{-\kappa})$
term parametrizes the correction due to the leading irrelevant
perturbation, see Eq.~\eqref{Hg3-scaling}. If
instead $\omega > 3/\nu-3$ then $\kappa=3/\nu-3$, the
$O(L^{-\kappa})$ term takes into account the analytic background in
Eq.~\eqref{Hg3-scaling}, and in this case $n_s=0$ should be used.

\begin{figure}
  \includegraphics[width=0.95\columnwidth]{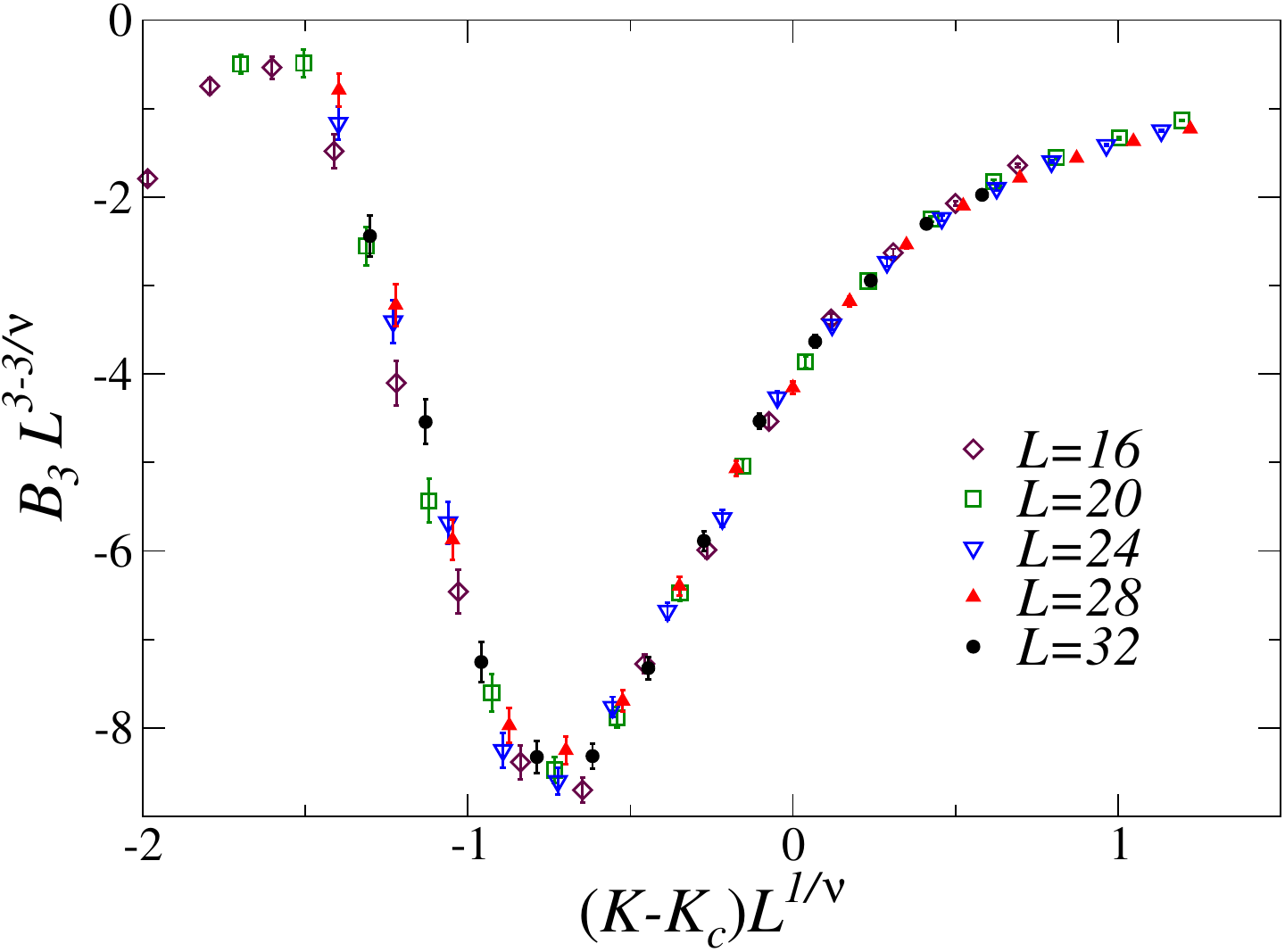}
  \caption{Scaling of third cumulant $B_3$ for $q=0.015$, obtained by plotting
    $L^{3-3/\nu} B_3$ versus $(K-K_c)L^{1/\nu}$ with our optimal
    estimates $K_c=0.8940$ and $\nu=0.82$.}
  \label{thirdcumsca}
\end{figure}

The outcomes of some of these
fits are reported in Table~\ref{TabFitRes}, and our optimal estimates are
\begin{equation}
  K_c = 0.8940(8),\qquad \nu = 0.82(2).
  \label{numest}
\end{equation}
The uncertainty on these estimates is dominated by the systematic
errors, which have been estimated by varying the degree of the
polynomial approximations (which has in fact a very mild effect), by
varying the fit range, by systematically discarding the data coming
from the smaller lattices in the fit to check the effects of scaling
corrections, and by adding some scaling corrections, see
Table~\ref{TabFitRes}. In Fig.~\ref{thirdcumsca} we show the resulting
FSS of $B_3$ when using our optimal estimates reported in
Eq.~(\ref{numest}), showing the good quality of the collapse of the
data.~\footnote{The estimate of $K_c$ that we report in
Eq.~(\ref{numest}) is significantly more accurate than the estimate
reported in Ref.~\cite{OAIM-04}, which was $K_c\approx 0.866$, somehow
obtained from the specific-heat data for $L=24$ at $q=0.015$.}

A reasonable hypothesis is that the value $\nu=0.82(2)$ of the
length-scale critical exponent characterizes the critical behavior
along the whole transition line up to to the Nishimori point. Note
that the value of $\nu$ that we have obtained is significantly larger
that the Ising exponent $\nu_{\cal I}\approx 0.630$ in the absence of
disorder. Moreover, the corresponding value of the specific-heat
exponent is negative, $\alpha=2-3\nu=-0.46(6)$, as expected by the
Harris criterium.  We also note that the value $\nu$ obtained by our
numerical analysis, cf. Eq.~(\ref{numest}), implies that the
background term in the FSS relation in Eq.~\eqref{Hg3-scaling} for the
third cumulant is suppressed as $O(L^{-(3/\nu-3)})$ with
$3/\nu-3=0.66(7)$. Further $O(L^{-\omega})$ scaling corrections
arising from the leading irrelevant RG perturbation are also expected,
however the available data are not sufficiently precise to reliably
estimate the value of $\omega$.

\begin{figure}
  \includegraphics[width=0.95\columnwidth]{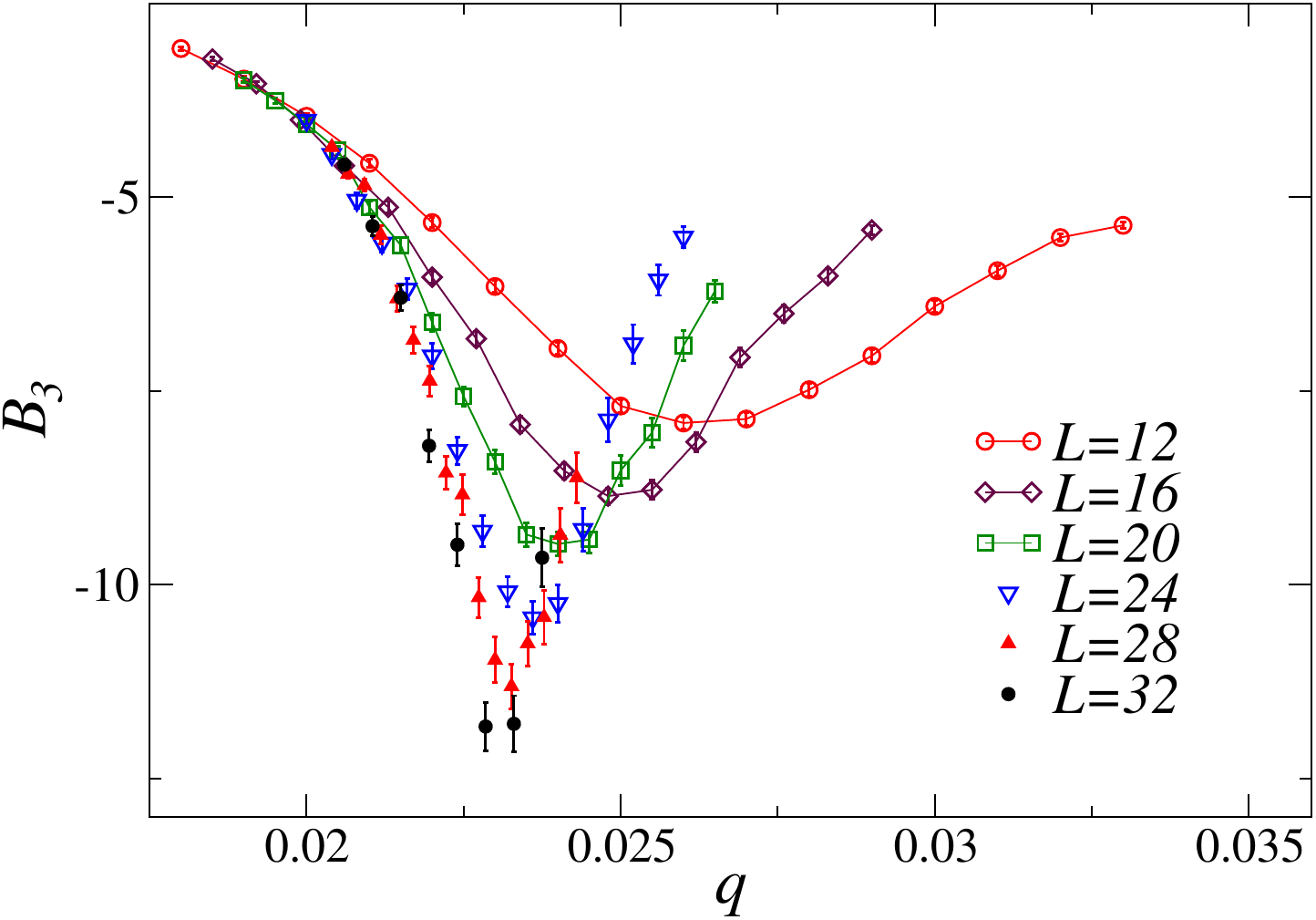}
  \caption{Data for the third cumulant $B_3$ for $K=1$, across
    the topological transition around $q\approx 0.022$. Continuous
    lines for $L=12, 16, 20$ are drawn just to guide the eye.}
  \label{thirdcum_K1}
\end{figure}

To verify that the universality class of the transitions does not
change along the confinement-deconfinement transition line for weak
disorder, we now present results obtained by fixing $K=1.0$,
performing a scan in $q$.  Also in this case the specific heat does
not display a divergent behavior, and the first seemingly divergent
cumulant is the third one, see Fig.~\ref{thirdcum_K1}. To extract the
critical properties from these data we can follow the same strategy
already adopted for the transition at $q=0.015$, the only difference
being that the scaling variable to be used in Eq.~\eqref{Hg3-scaling}
is now
\begin{equation}\label{Xdefbis}
X=(q-q_c)L^{1/\nu}
\end{equation}
instead of the definition reported in Eq.~\eqref{Hg3-scaling}. The
final result of this analysis are the estimates
\begin{equation}\label{numestbis}
q_c=0.0219(3),\qquad \nu=0.84(8), 
\end{equation}
where the reported errors are largely dominated by systematic errors.
The estimated value of the critical exponent $\nu$ is fully consistent
with the one obtained previously, see Eq.~\eqref{numest}, although
significantly less accurate.

\begin{figure}
  \includegraphics[width=0.95\columnwidth]{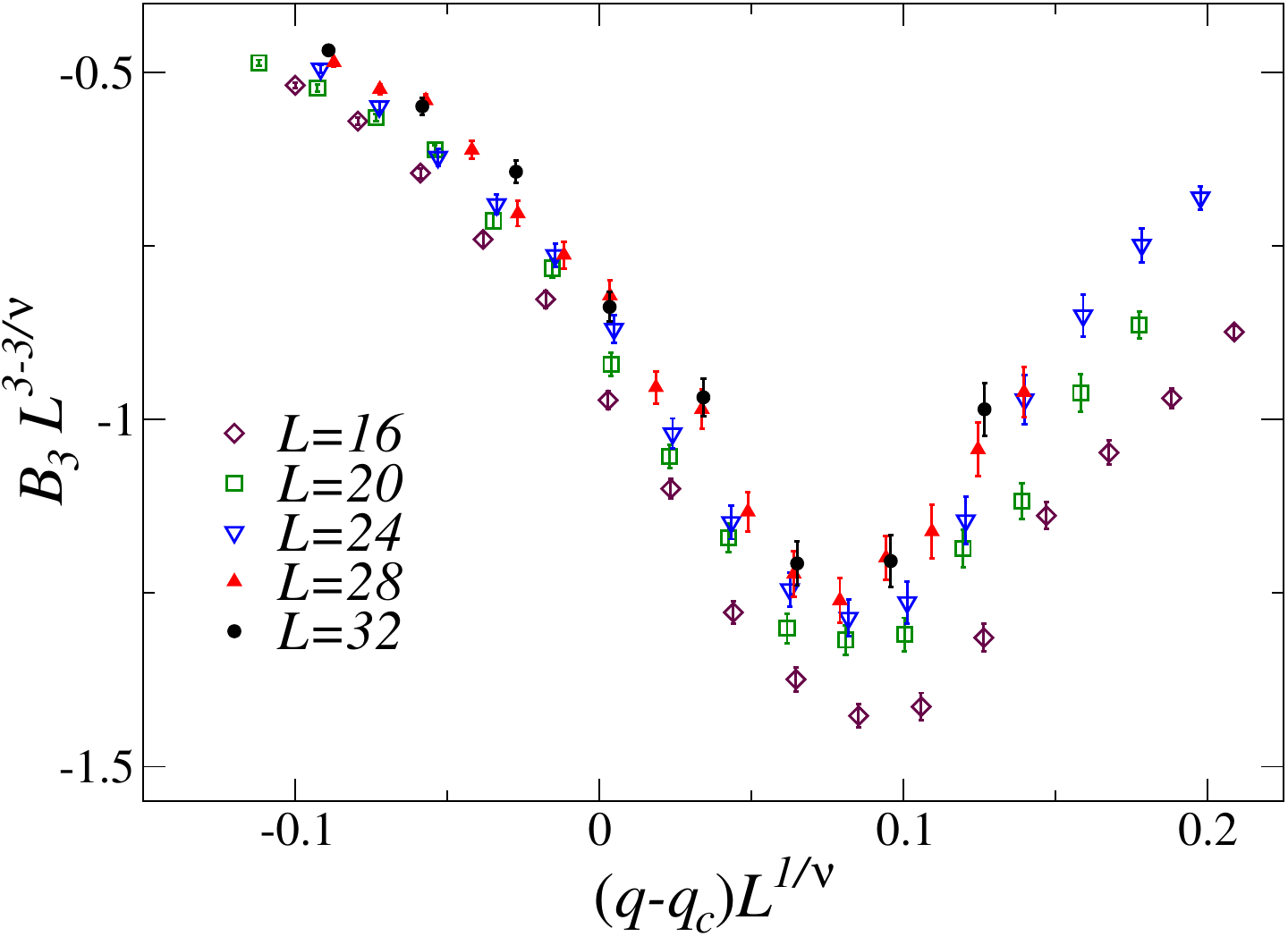}
  \caption{Scaling of third cumulant $B_3$ for $K=1$,
      obtained by plotting $L^{3-3/\nu} B_3$ versus $(q-q_c)L^{1/\nu}$
      using our optimal estimates $q_c=0.0219$ and the value
      $\nu=0.82$ obtained in the $q=0.015$ case.}
  \label{thirdcumsca_K1}
\end{figure}

In Fig.~\ref{thirdcumsca_K1} we show the scaling plot obtained by
rescaling $B_3$ data using the optimal value $q_c$ in
Eq.~\eqref{numestbis} and the critical exponent $\nu=0.82$ obtained at
fixed $q=0.015$. The results are substantially consistent, however we
note that scaling corrections are much larger than those at fixed
$q=0.015$, see Fig.~\ref{thirdcumsca}. This slower approach
  to the asymptotic critical behavior can be easily explained by the
  presence of larger scaling corrections, as generally expected when
  approaching the multicritical Nishimori point, where the
critical behavior is expected to change.  The qualitative
behaviors of the $B_3$ curves obtained for $K=1$ and $q=0.015$ is the
same, however the presence of large scaling corrections for $K=1$
makes it impossible to achieve a robust quantitative check that the
asymptotic scaling curves associated with $B_3$, cf.
Eq.~\eqref{Hg3-scaling}, are the same up to rescaling coefficients in
the two cases.

In conclusion, we believe that our results provide a quite robust
evidence that the topological transitions of the 3D lattice ${\mathbb
  Z}_2$ gauge model changes universality class in the presence of
quenched uncorrelated disorder associated with the
plaquette. Moreover, we note that the exponent $\nu$ of this new
universality class significantly differs from that of the
ferromagnetic transitions in 3D random-bond Ising models, for
which~\cite{HPPV-07} $\nu_{\rm rdi}=0.683(2)$.

\section{Conclusions}
\label{conclu}

We have investigated the effects of quenched disorder at
finite-temperature topological transitions, focusing on the 3D
${\mathbb Z}_2$ gauge model in the presence of an uncorrelated
quenched disorder associated with the plaquettes of the lattice
(RPGM).  This is a paradigmatic model for systems undergoing
finite-temperature topological transitions~\cite{Wegner-71, Kogut-79,
  Sachdev-19, BPV-25}, indeed its confinement-deconfinement
transitions are not driven by any local order parameter. We report a
numerical analysis of their critical behaviors for weak disorder, and
for this purpose we focus on the scaling behavior of the
gauge-invariant energy cumulants defined in Eq.~(\ref{bkdef}).  The
numerical investigation has been carried out using the binomial
distribution for disorder variables, but it is natural to
expect that the universality class of the results is
independent of the disorder distribution, as it happens for the Ising
model with quenched disorder, see, e.g., Ref.~\cite{HPPV-07-rdi}.

We show that the topological transitions of the 3D RPGM belong to a
new topological universality class, different from that of the pure
system, in agreement with the Harris criterium for the relevance of
uncorrelated quenched disorder when the pure system undergoes a
continuous transition with positive specific-heat critical
exponent. Indeed, the critical behavior in the presence of weak
disorder turns out to be characterized by the length-scale critical
exponent $\nu = 0.82(2)$, which is significantly larger that the Ising
length-scale critical exponent $\nu_{\cal I}\approx 0.630$ in the
absence of disorder.

The same universality class is expected to describe all the
transitions happening for $q<q_N$ and $K<K_N$, where $(q_N, K_N)$ is
the intersection point of the deconfinement transition line and the
Nishimori line, see Fig.~\ref{phadia}.  At this intersection point the
RPGM is expected to develop a peculiar multicritical behavior,
characterized by two relevant RG perturbations with RG dimensions
$y_1>y_2>0$.  Analogously to the multicritical behavior at the
Nishimori point of the 2D and 3D random-bond Ising
models~\cite{Nishimori-81,LH-88,HPPV-07-mc,HPPV-08}, we expect that
the multicritical behavior of RPGM is characterized by two relevant RG
perturbations associated with the scaling fields $u_1=q-q_N$, which
runs parallel to the transition line at the Nishimori point, and $u_2=
{\rm tanh} K + 2q - 1$, with corresponding RG dimensions $y_1$ and
$y_2$ respectively, so that $y_1>y_2>0$.  Unfortunately, this
multicritical behavior cannot be investigated using the techniques
adopted in the present study. This is due to the fact that the energy
density, and all the energy cumulants, along the Nishimori line are
analytic functions, indeed the exact result~\cite{WHP-03}
\begin{equation}
  \frac{1}{V} [\langle \mathcal{H}\rangle]_w =3-6q
  \label{enenline}
\end{equation}
holds along the Nishimori line (\ref{nishimoriline}) (see also
analogous behaviors~\cite{Nishimori-book, Nishimori-81} for the
random-bond Ising model).  Therefore, the scaling part of the
cumulants vanishes along the Nishimori line, therefore to study the
singular behavior of the free energy one cannot move along the
Nishimori line. This makes an investigation of the multicritical
behavior at the Nishimori point extremely challenging from the
computational point of view, since it requires the independent tuning
of two different parameters (i.e., both $K$ and $q$) to approach the
unknown position of the multicritical point. We also note that the
critical behaviors along the deconfinement-confinement transition line
below the Nishimori point, see Fig.~\ref{phadia}, is expected to
belong to a different universality class, as it occurs for other
systems with quenched disorder, see, e.g.,
Refs.~\cite{CPV-11,AV-11,PPV-09}.

We finally discuss possible extensions of this study to generic
lattice ${\mathbb Z}_N$ gauge models with $N>2$ in the presence of
quenched disorder associated with the plaquette, which are simply
obtained by replacing the ${\mathbb Z}_2$ link variable $\sigma_{{\bm
    x},\nu}$ in the plaquette operator (\ref{plaquette}) with
$\lambda_{{\bm x},\nu} = \exp(i 2\pi n_{{\bm x},\nu}/N)$ with $n_{{\bm
    x},\nu} = 0,1,...,N-1$, and using the real part of the plaquette
operator in the Hamiltonian (for $N=2$ we recover the ${\mathbb Z}_2$
gauge model).  We recall that, in the absence of quenched disorder,
therefore for $q=0$ in Eq.~(\ref{probdis}), 3D lattice ${\mathbb Z}_N$
gauge models undergo finite-temperature topological transitions for
any $N$. For $N>4$ the transitions are continuous and belong to the
inverted XY (IXY) universality class~\cite{SSNHS-03, BCCGPS-14,
  BPV-22-dis,BPV-25}. For $N=4$ the transition is continuous and
belongs to the Ising universality class because its partition
functions can be written as the square of the partition function of
the lattice ${\mathbb Z}_2$ gauge model (note however that this holds
only for the specific plaquette Hamiltonian that we
consider~\cite{BPV-25}). Finally the transition is first order for
$N=3$. Therefore, since the specific-heat critical exponent of the IXY
universality class is negative~\cite{PV-02,BPV-25} (it is equal to
that of the duality related XY universality class, $\alpha\approx
-0.015$), according to the Harris criterium we expect that a weak
quenched disorder is irrelevant for $N>4$, and the topological
transitions still belong to the IXY universality class. On the other
hand, the quenched disorder must be relevant for $N=4$ if the
plaquette action is used, because its specific heat exponent is
positive. Indeed, analogously to the pure
case~\cite{BPV-24-coH,Korthals-78}, one can easily prove that for each
disorder configuration the partition function of $N=4$ model satisfies
the identity
\begin{equation}
  Z_{N=4,w}(K) = Z_{N=2,w}(K/2)^2.
  \label{zn4}
 \end{equation}
The corresponding free-energy density (\ref{Fkp}) in the presence of
disorder is thus equivalent to that of the 3D lattice ${\mathbb Z}_2$
gauge model. Therefore, we expect to observe continuous transitions
analogous to those found in the 3D lattice ${\mathbb Z}_2$ gauge model
with quenched disorder, apart from a trivial change of the coupling
$K$ according to the identity (\ref{zn4}). On the other hand, the
situation is less clear for $N=3$: the first-order transitions of pure
systems are generically expected to be smoothed by the quenched
disorder (see, e.g., \cite{Cardy-99}), and may give rise to continuous
transitions belonging to a new universality class.

\acknowledgments

Numerical simulations have been performed on the CSN4 cluster of the Scientific
Computing Center at INFN-PISA.

\end{document}